\documentclass[prb,twocolumn,showpacs,preprintnumbers,amsmath,amssymb,floatfix]{revtex4}
\usepackage[dvips]{graphics,graphicx,color}

\begin{document}

\title{Van der Waals Interactions Between Thin Metallic Wires and Layers}

\author{N.\ D.\ Drummond and R.\ J.\ Needs}

\affiliation{TCM Group, Cavendish Laboratory, University of Cambridge,
Cambridge CB3 0HE, United Kingdom}

\begin{abstract}
Quantum Monte Carlo (QMC) methods have been used to obtain accurate
binding-energy data for pairs of parallel, thin, metallic wires and layers
modelled by 1D and 2D homogeneous electron gases.  We compare our QMC binding
energies with results obtained within the random phase approximation, finding
significant quantitative differences and disagreement over the
asymptotic behavior for bilayers at low densities.  We have calculated
pair-correlation functions for metallic biwire and bilayer systems.  Our QMC
data could be used to investigate van der Waals energy functionals.
\end{abstract}


\maketitle

One-dimensional conductors such as carbon nanotubes are essential
components of many proposed nanotechnological devices, and are currently the
subject of numerous experimental and theoretical studies.  It has recently
been demonstrated\cite{dobson} that the van der Waals (vdW) interaction
between pairs of distant, parallel, thin conducting wires assumed in many
current models of metallic carbon nanotubes is qualitatively wrong.  In this
letter we provide the first accurate binding-energy data for pairs of thin
metallic wires and layers, which can be used as a benchmark for subsequent
theoretical studies or to parametrize model interactions between 1D
and 2D conductors.

Thin, electrically neutral wires are attracted to one another by vdW forces.
The standard method for calculating the vdW interaction between objects is
to assume there are pairwise interactions between volume elements with an
attractive tail of the form $U_{\rm PP}(r) \propto -r^{-6}$, which is
appropriate for the vdW interaction between molecules.  Summing these
interactions for a pair of 1D parallel wires separated by a distance $z$ gives
a vdW binding energy of $U(z) \propto -z^{-5}$.  Such pairwise vdW models have
been used in studies of single-walled carbon nanotubes.\cite{girifalco,sun}
However, a recent investigation\cite{dobson} of the interaction between pairs
of thin, metallic wires modelled by 1D homogeneous electron gases (HEGs)
within the random phase approximation (RPA) found that the binding energy
falls off (approximately) as\cite{hartree_au}
\begin{equation} 
U(z) \approx -\frac{\sqrt{r_s}}{16 \pi z^2 \left[ \log(2.39z/b)
\right]^{3/2}}, \label{eqn:rpa_biwire_binding}
\end{equation} 
where $b$ is the wire radius and $2r_s$ is the length of the wire section
containing one electron on average.  The pairwise vdW model is clearly
appropriate for an insulator or for a metallic wire whose radius is greater
than the screening length, but is inappropriate for a thin conductor such as a
single-walled carbon nanotube.\cite{dobson}

Likewise, the binding energy per particle of a pair of thin, parallel metallic
layers can be shown to decay as
\begin{equation} 
U(z)=\frac{-0.012562 \sqrt{\pi} r_s}{2 z^{5/2}}
\label{eqn:rpa_bilayer_binding}
\end{equation} 
within the RPA,\cite{sernelius,dobson} compared with $U(z) \propto -z^{-4}$
within the pairwise vdW theory, where $z$ is the layer separation.  (In a 2D
HEG $r_s$ is the radius of the circle that contains one electron on average.)
At very large separations the vdW attraction is dominated by the Casimir
effect,\cite{casimir,sernelius} in which the zero-point energy of photon modes
between the metallic layers gives rise to an attractive force. However we
restrict our attention to the range of separations in which vdW effects are
dominant.

Within the RPA, the binding energy may be calculated as the change in the
zero-point energy of plasmon modes as a function of separation.\cite{dobson}
However, the RPA is poor in low-dimensional systems, and ceases to be valid at
low densities, where correlation effects become dominant.  We have therefore
performed quantum Monte Carlo\cite{ceperley_1980,foulkes_2001} (QMC)
calculations of the binding energies of pairs of thin, metallic wires and
layers modelled by 1D and 2D HEGs with neutralizing backgrounds.  In
particular we have used the variational and diffusion quantum Monte Carlo (VMC
and DMC) methods as implemented in the \textsc{casino} code.\cite{casino} DMC
is the most accurate method available for studying quantum many-body systems
such as electron gases.  We have also calculated pair-correlation functions
(PCFs), enabling us to examine the correlation hole responsible for the vdW
attraction between pairs of wires and layers.

In our QMC calculations we use the full Coulomb potential, so that for 1D HEGs
the many-electron wave function must go to zero at both parallel- and
antiparallel-spin coalescence points for electrons in the same wire.  The
nodal surfaces for paramagnetic and ferromagnetic 1D HEGs are therefore the
same, and so the fixed-node DMC energy---which is equal to the exact
ground-state energy because the nodal surface is exact---is independent of the
spin polarization.  This conclusion does not violate the Lieb-Mattis
theorem\cite{lieb} because the 1D Coulomb interaction is pathological in the
formal sense of Lieb and Mattis.\cite{lm_comment} For convenience we
choose to work with ferromagnetic 1D HEGs, because a Slater determinant wave
function produces the correct nodal surface in this case.  Our QMC studies of
1D HEGs will be published elsewhere.\cite{ndd_unpublished}

Fermionic symmetry is imposed in 2D via the fixed-node
approximation,\cite{anderson_1976} in which the nodal surface is constrained
to equal that of a trial wave function.  We expect a very high degree of
cancellation of fixed-node errors when the binding energy is calculated.  It
has already been shown that fixed-node DMC is able to describe vdW forces
between helium\cite{helium_vdW_QMC} and neon atoms.\cite{ndd_neon}

We have verified that time-step and population-control biases in our DMC
energies are negligible by repeating some of the calculations using different
time steps and populations.  Finite-size errors are a more serious problem,
although most of the bias cancels out when the energy difference is taken to
obtain the binding energy.  Twist averaging\cite{lin_2001} or the addition of
finite-size corrections are unlikely to reduce the bias in the binding energy.
We expect our binding-energy results to be valid so long as the wire or layer
separation is small compared with the length of the 1D or 2D simulation cell;
for larger separations the system resembles a pair of insulators and the
binding energy is expected to fall off more steeply (in accordance with the
pairwise vdW model).

We have used Slater-Jastrow-backflow trial wave
functions.\cite{foulkes_2001,backflow} For our 2D bilayer calculations the
Slater part of the wave function consists of a product of four determinants of
plane-wave orbitals for spin-up and spin-down electrons in each of the two
layers.  For our 1D biwire calculations the Slater part consists of a product
of two determinants of plane-wave orbitals for the electrons in each wire
(recall that each wire is ferromagnetic in our calculations).  Slater
determinants for a 1D HEG are of Vandermonde form, and could therefore be
rewritten as polynomials and evaluated in a time that scales linearly with
system size; however other parts of the QMC algorithm such as the evaluation
of the two-body Jastrow terms and backflow functions take up a significant
fraction of the computer time, so for convenience we have continued to employ
the usual determinant-evaluating and updating machinery of QMC
calculations.\cite{fahy_1990}

Our Jastrow factors consist of polynomial and plane-wave two-body
terms\cite{ndd_jastrow} satisfying the Kato cusp conditions.\cite{kato_pack}
In spite of the fact that the nodal surface is exact in 1D, two-body backflow
correlations\cite{backflow} were found to make a very significant improvement
to the wave function, as can be seen in Table \ref{table:wf_quality}.
Backflow functions were used in all of our 1D calculations and in our 2D
calculations at $r_s=1$~a.u.  Free parameters in the trial wave function were
optimized by minimizing the unreweighted variance of the local
energy.\cite{umrigar_1988a,ndd_newopt}

\begin{table}
\begin{center}
\begin{tabular}{lr@{.}lr@{.}lr@{}l}
\hline \hline Method & \multicolumn{2}{c}{Energy (a.u.\ / el.)} &
\multicolumn{2}{c}{Var.\ (a.u.)} & \multicolumn{2}{c}{\%age corr.\ en.} \\
\hline

HF      & ~$-0$&$215943040112$ & \multicolumn{2}{c}{---} &
\hspace{2em}$0$&\% \\

SJ-VMC  & $-0$&$2319668(4)$   & ~$0$&$0000360(2)$ & $99$&.$974(3)$\% \\

SJB-VMC & $-0$&$2319710(3)$   & $0$&$0000046(1)$ & $100$&.$000(3)$\% \\

SJB-DMC & $-0$&$2319709(3)$   & \multicolumn{2}{c}{---} & $100$&\% \\

\hline \hline
\end{tabular}
\end{center}
\caption{Energy, energy variance, and fraction of correlation energy retrieved
  using different levels of theory and wave function for a 15-electron 1D
  ferromagnetic HEG at $r_s$=15~a.u.  ``HF'' stands for Hartree-Fock theory,
  ``SJ'' denotes a Slater-Jastrow trial wave function and ``SJB'' a
  Slater-Jastrow-backflow trial wave function.
\label{table:wf_quality}}
\end{table}

Each wire or layer is accompanied by a neutralizing background.  When studying
biwires (bilayers) we need to add the interaction between the electrons in
each wire (layer) and the background of the opposite wire (layer), plus the
interaction of the two backgrounds.  This contribution to the energy is
$E_{\rm cap}=\log(z)/(2r_s)$ for a biwire and $E_{\rm cap}=z/r_s^2$ for a
bilayer.

\begin{figure}
\begin{center}
\includegraphics[scale=0.33,clip]{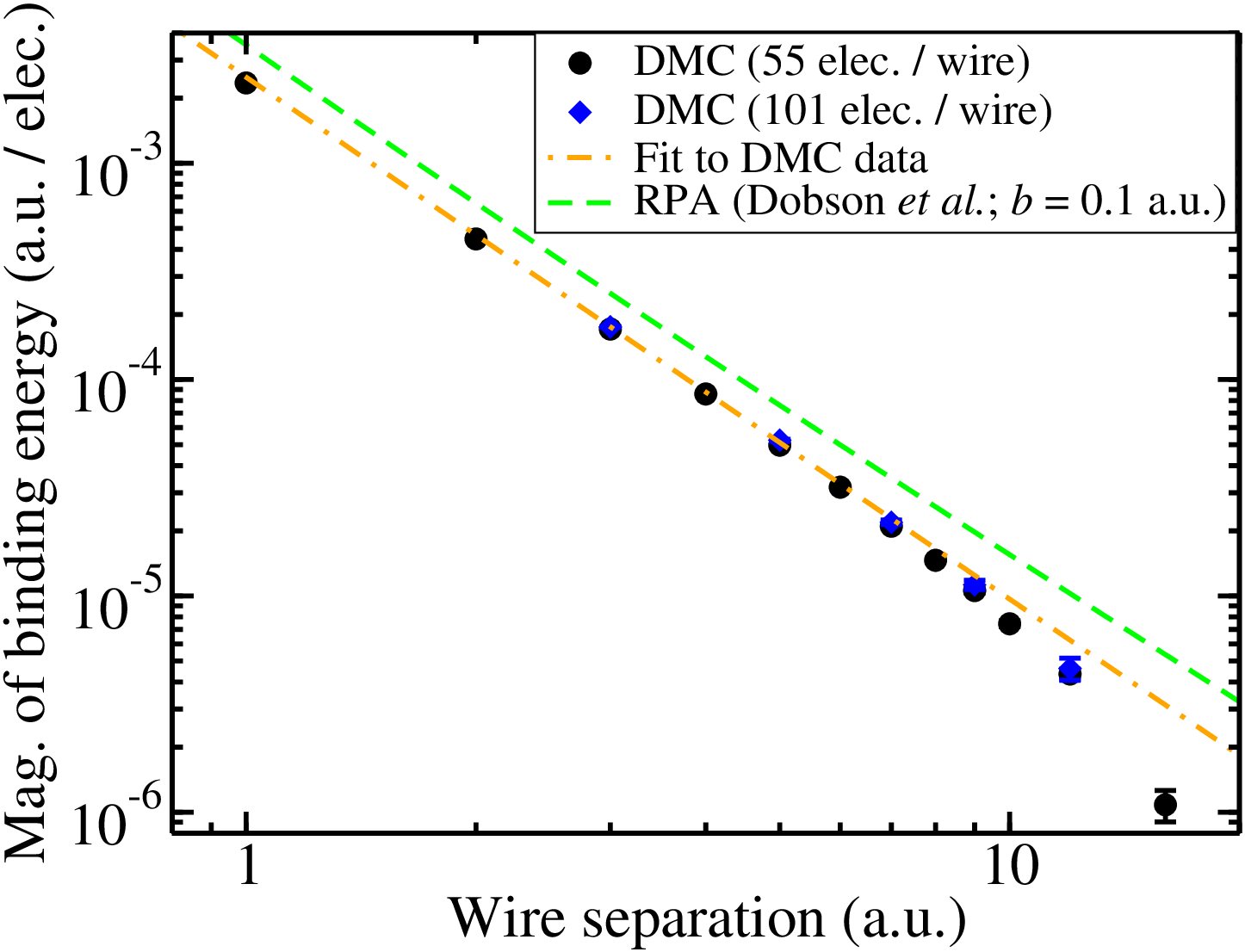} \\
\includegraphics[scale=0.33,clip]{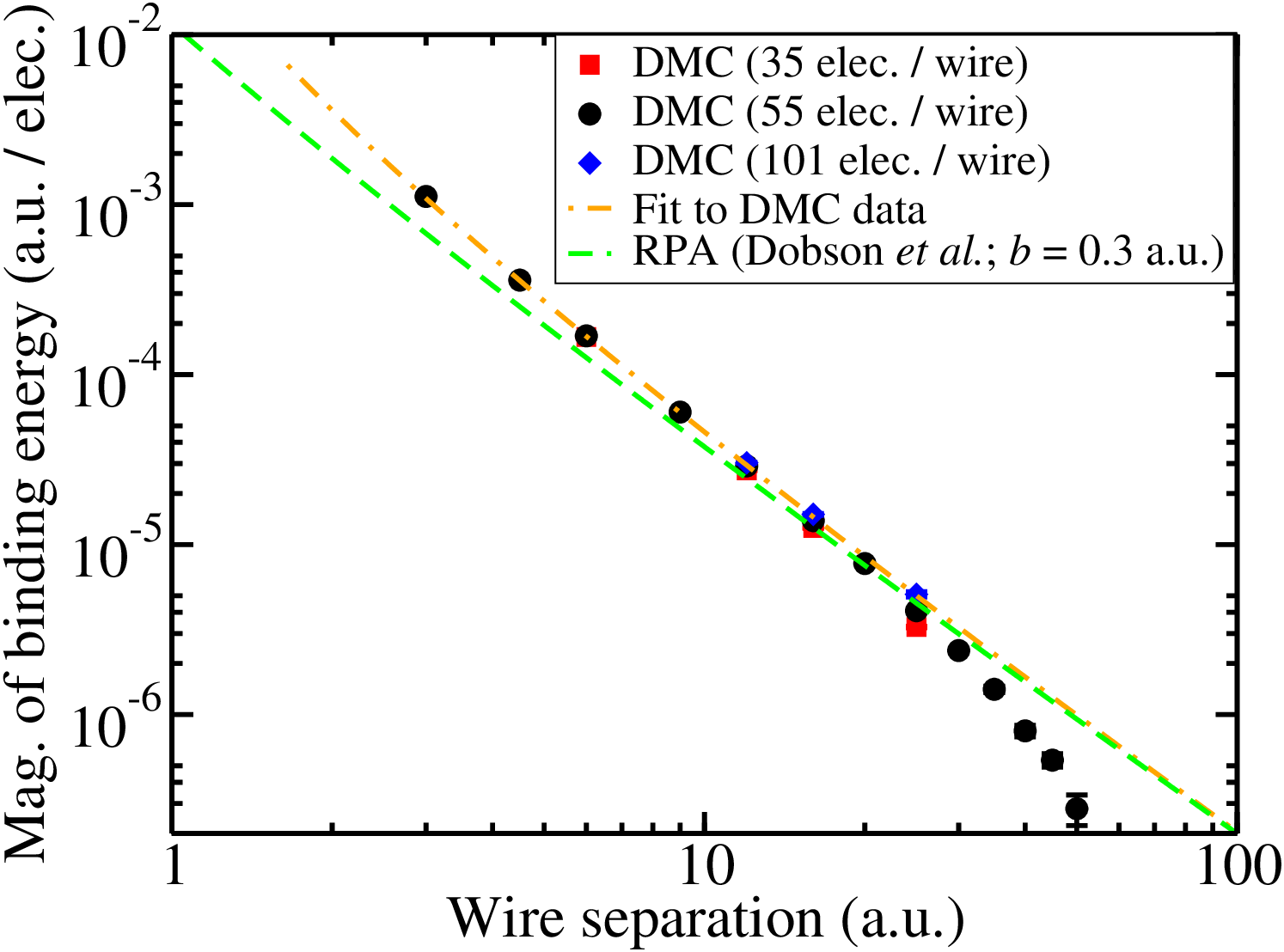} \\
\includegraphics[scale=0.33,clip]{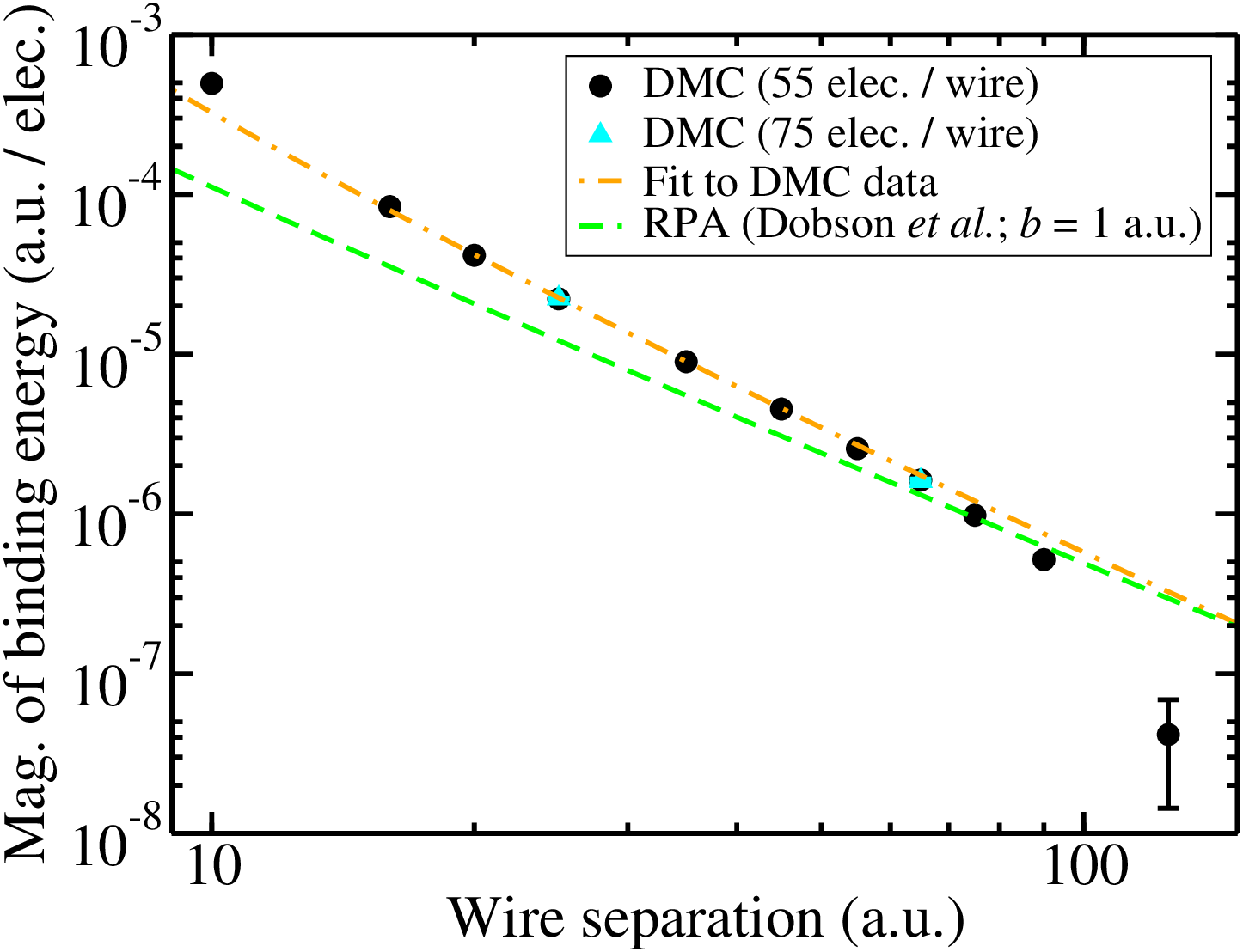}
\caption{(Color online) Binding energy per particle of a 1D HEG biwire as a
  function of wire separation for $r_s=1$~a.u.\ (top panel), $r_s=3$~a.u.\
  (middle panel), and $r_s=10$~a.u.\ (bottom panel).  The DMC time steps were
  0.04, 0.2, and 2.5~a.u.\ at $r_s=1$, 3, and 10~a.u., and the target
  configuration population was 2048 in each case.
\label{figure:biwire_binding}}
\end{center}
\end{figure}

\begin{figure}
\begin{center}
\includegraphics[scale=0.33,clip]{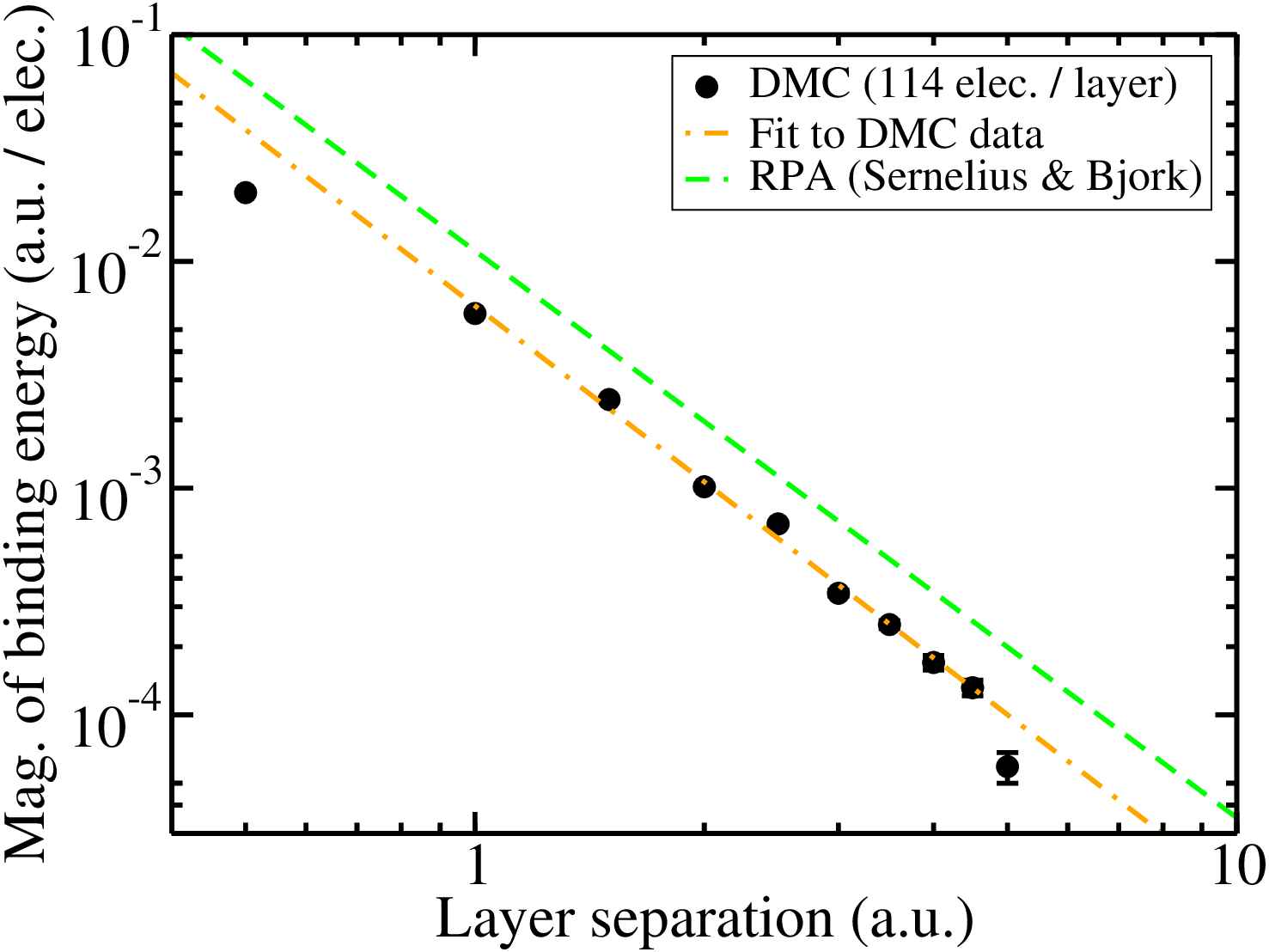} \\
\includegraphics[scale=0.33,clip]{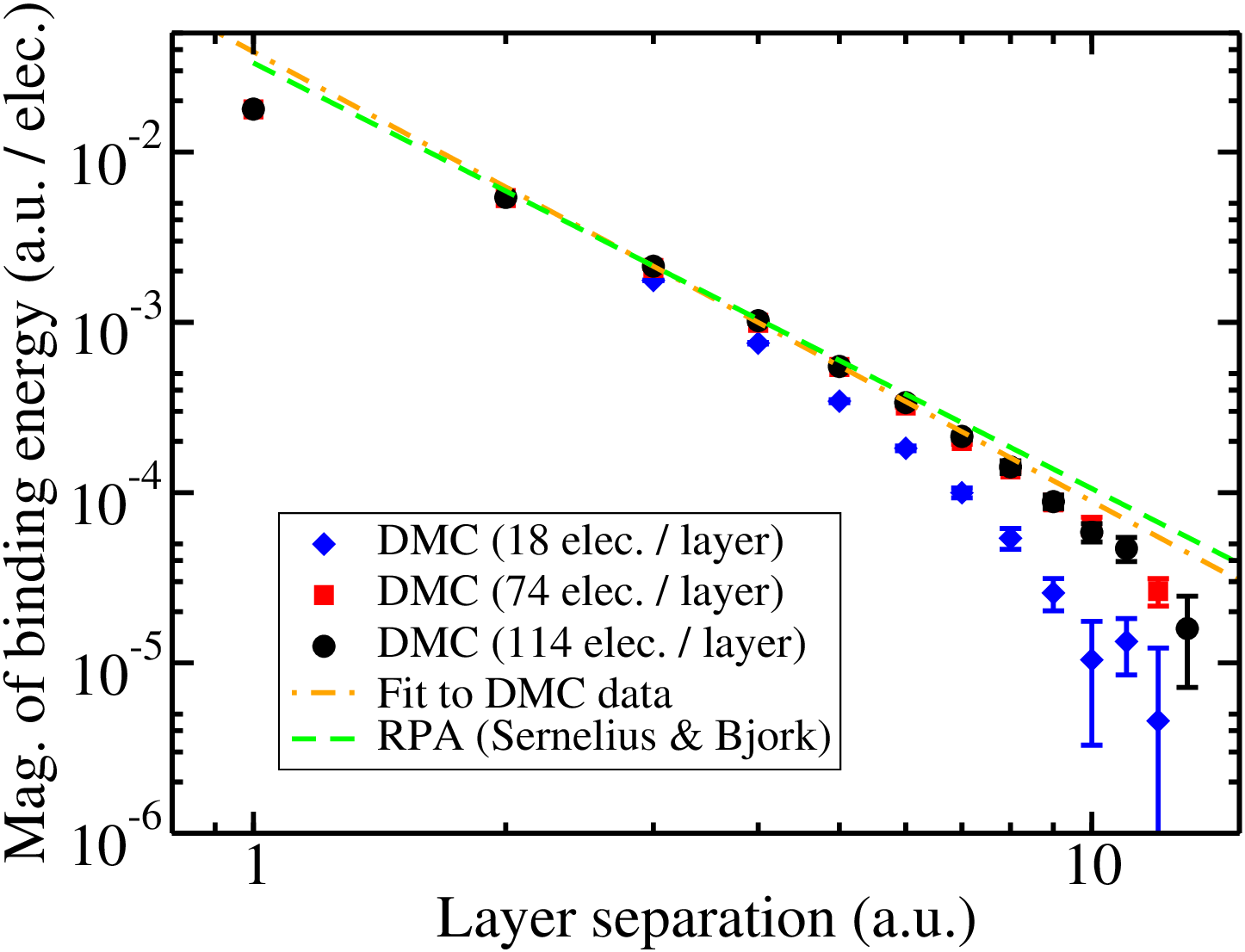} \\
\includegraphics[scale=0.33,clip]{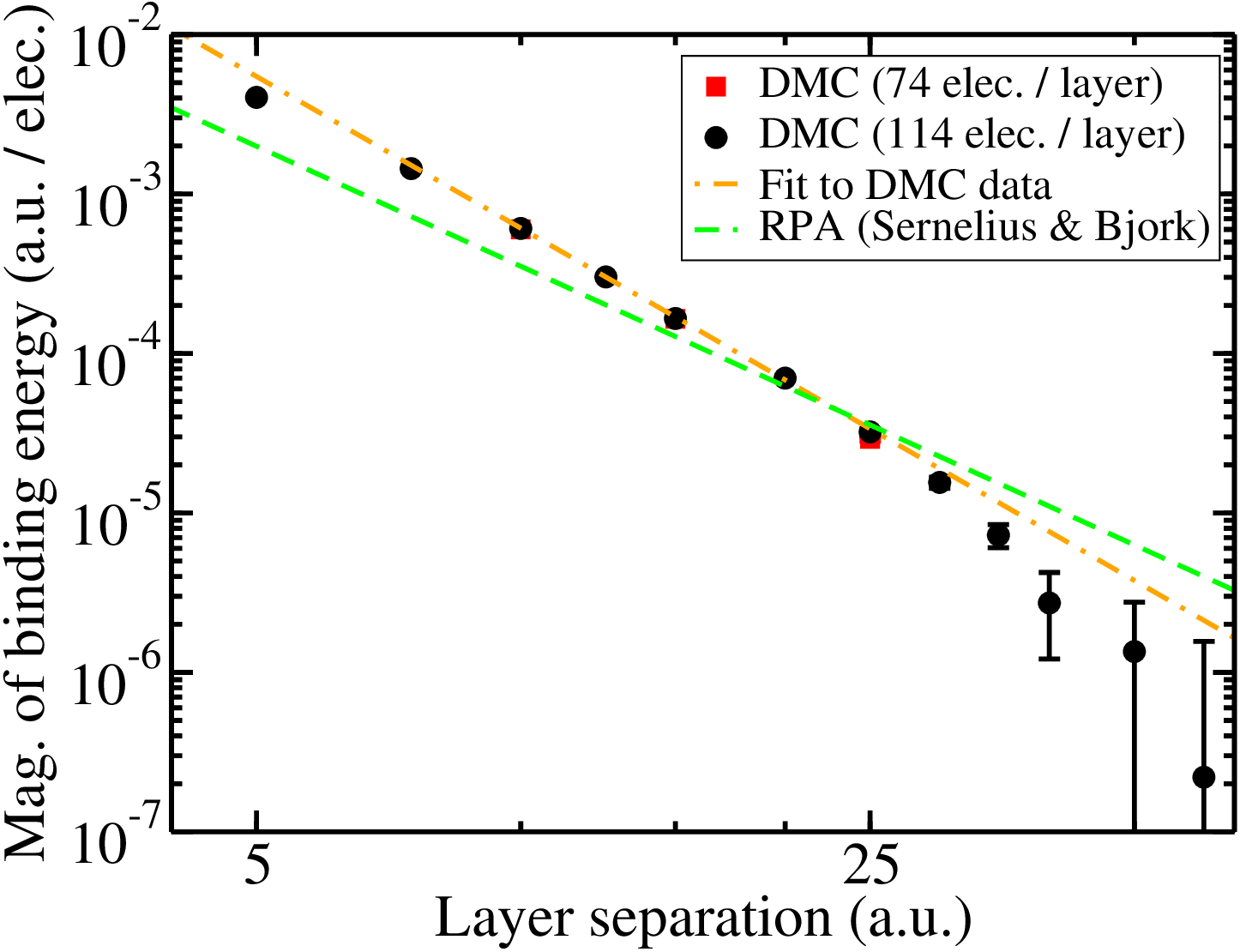}
\caption{(Color online) Binding energy per particle of a 2D HEG bilayer as a
  function of layer separation for $r_s=1$~a.u.\ (top panel), $r_s=3$~a.u.\
  (middle panel), and $r_s=10$~a.u.\ (bottom panel). The DMC time steps were
  0.007, 0.05, and 0.5~a.u.\ at $r_s=1$, 3, and 10~a.u., and the target
  configuration populations were 1024, 320, and 1024.
\label{figure:bilayer_binding}}
\end{center}
\end{figure}

The binding energies of pairs of 1D and 2D HEGs are shown in Figs.\
\ref{figure:biwire_binding} and \ref{figure:bilayer_binding}.  The approximate
RPA binding energy shown in Fig.~\ref{figure:biwire_binding} was obtained
using $b=r_s/10$ in Eq.~(\ref{eqn:rpa_biwire_binding}).  The wire radius $b$
is therefore small compared with the other length scales in the system.  We
have included DMC results at several different system sizes in Figs.\
\ref{figure:biwire_binding} and \ref{figure:bilayer_binding}.  In 1D, half the
length of the simulation cell is $L_{r_s,N}=Nr_s$, where $N$ is the number of
electrons per wire, and in 2D the size of the simulation cell is $L_{r_s,N}
\approx \sqrt{N}r_s$.  The binding energy falls off more steeply once $z$
becomes a significant fraction of $L_{r_s,N}$, as expected.  Clearly the
binding energies enter the asymptotic regime when $z \gg r_s$.  We have
therefore fitted the RPA asymptotic binding-energy forms to our QMC data in
the range $r_s \ll z \ll L_{r_s,N}$.  We believe the errors in the fitted
exponents are about $0.1$--$0.2$.

The fits to the DMC biwire binding-energy data shown in
Fig.~\ref{figure:biwire_binding} are
\begin{eqnarray}
U_1(z) & = & -0.0815 z^{-2.28} \left[ \log(27000z) \right]^{-3/2} \\ U_3(z) &
= & -0.0225 z^{-1.98} \left[ \log(1.95z) \right]^{-3/2} \\ U_{10}(z) & = &
-0.0967z^{-2.17} \left[ \log(0.492z) \right]^{-3/2},
\end{eqnarray}
where $U_{r_s}(z)$ is the binding energy at density parameter $r_s$.  The DMC
binding-energy data are clearly in much better agreement with the RPA
[Eq.~(\ref{eqn:rpa_biwire_binding})] than with the pairwise vdW theory [$U(z)
\propto z^{-5}$].  It is not meaningful to compare the prefactors because of
the arbitrariness of our choice of the wire radius $b$ in the RPA theory.

The fits to the DMC bilayer binding-energy data shown in
Fig.~\ref{figure:bilayer_binding} are
\begin{eqnarray}
U_1(z) & = & -0.00637 z^{-2.58} \\  U_3(z) & = & -0.0388 z^{-2.64} \\
U_{10}(z) & = & -0.882 z^{-3.16},
\end{eqnarray}
where $U_{r_s}(z)$ is the binding energy at density parameter $r_s$.  At high
densities ($r_s=1$ and 3~a.u.)\ our results clearly show the $-z^{-2.5}$
behavior predicted by Eq.~(\ref{eqn:rpa_bilayer_binding}).  At low density
($r_s=10$~a.u.)\ the binding energy falls off more steeply than predicted by
the RPA, although the asymptotic behavior is clearly better described by the
RPA than the pairwise vdW theory [$U(z) \propto -z^{-4}$].  DMC and the RPA
give similar prefactors for the asymptotic binding energy at $r_s=3$~a.u., but
the DMC prefactor is somewhat lower at $r_s=1$~a.u.

\begin{figure}
\begin{center}
\includegraphics[scale=0.33,clip]{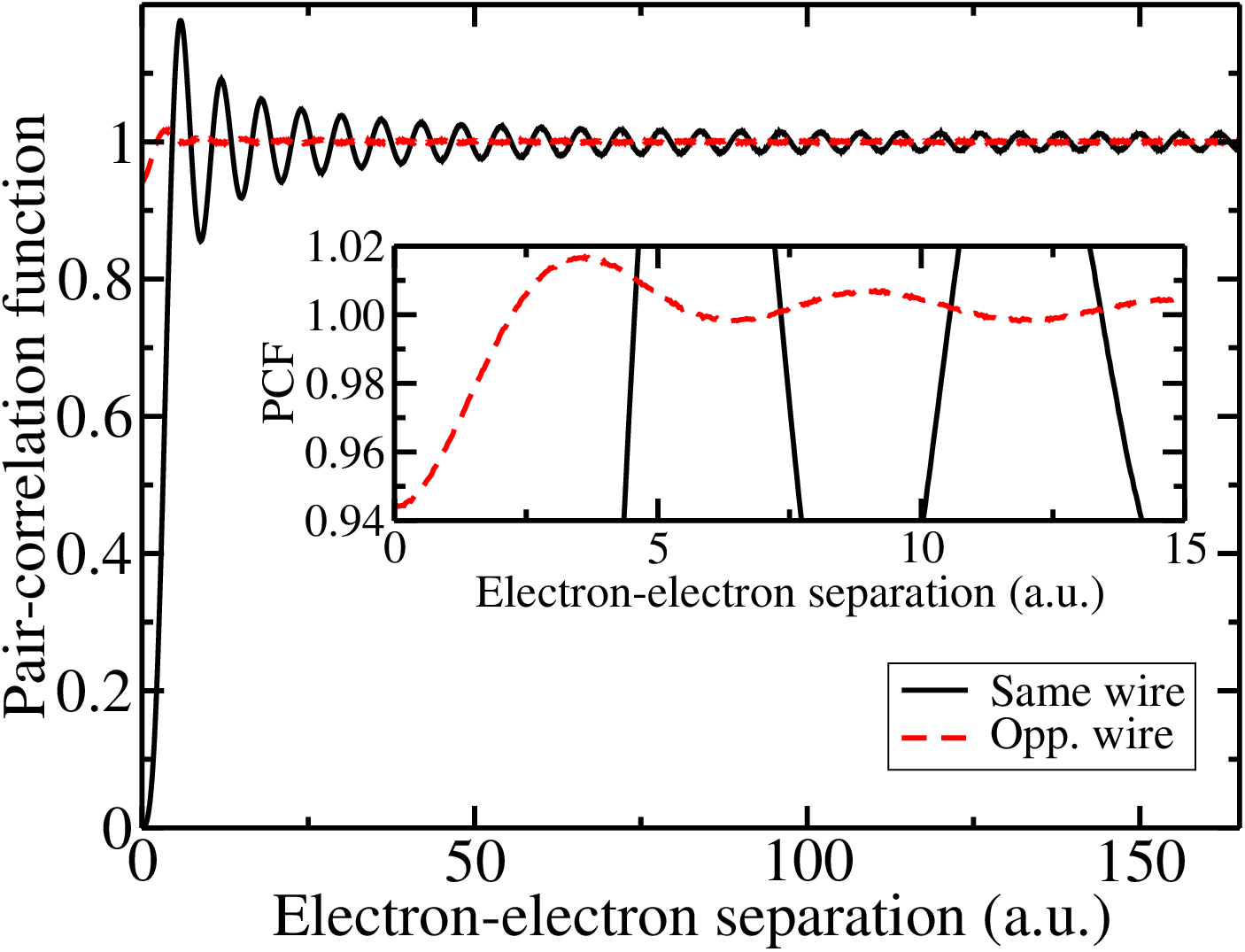}
\caption{(Color online) VMC PCF for a 1D HEG biwire at $r_s=3$~a.u.\ and wire
  separation $z=3$~a.u.  The inset shows the interwire correlation hole in
  greater detail.  It was verified for smaller system sizes that the DMC and
  VMC PCFs were in excellent agreement (as expected due to the accuracy of
  the trial wave function illustrated in Table \ref{table:wf_quality}).
\label{figure:biwire_pcf_rs3_z3}}
\end{center}
\end{figure}

\begin{figure}
\begin{center}
\includegraphics[scale=0.33,clip]{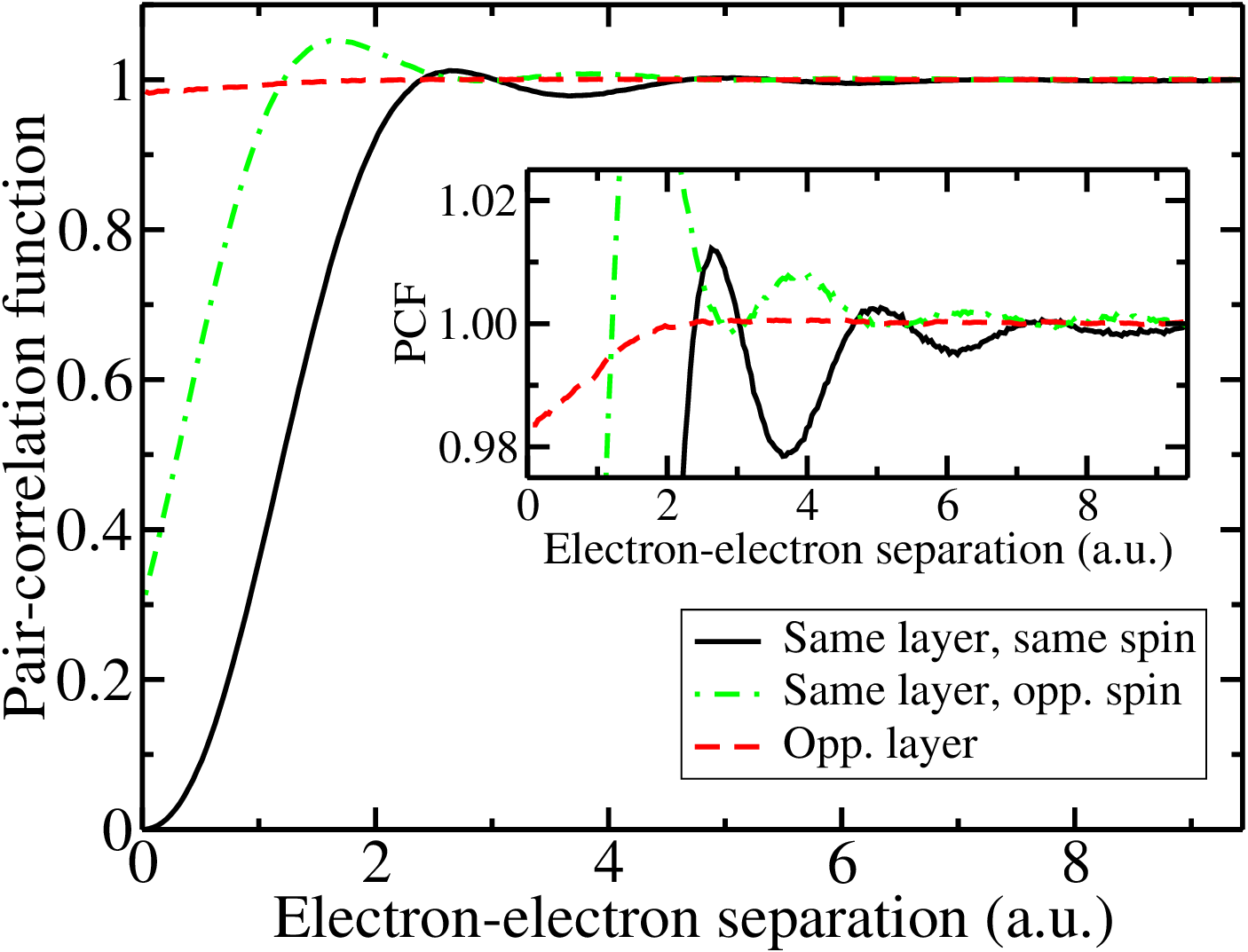}
\caption{(Color online) Extrapolated PCF for a 2D HEG bilayer at $r_s=1$~a.u.\
  and layer separation $z=2$~a.u.  The inset shows the interwire correlation
  hole in greater detail.  The VMC and DMC PCFs are in excellent agreement,
  implying that the extrapolated PCF is reliable.
\label{figure:bilayer_pcf_rs1_z2}}
\end{center}
\end{figure}

PCFs were accumulated by binning the interparticle distances in the electron
configurations generated by the VMC and DMC algorithms.  The error in the VMC
and DMC PCFs $g_{\rm VMC}$ and $g_{\rm DMC}$ is first order in the error in
the trial wave function, but the error in the extrapolated PCF $g_{\rm
ext}=2g_{\rm DMC}-g_{\rm VMC}$ is second order in the error in the wave
function.\cite{foulkes_2001} PCFs for biwires and bilayers are shown in Figs.\
\ref{figure:biwire_pcf_rs3_z3} and \ref{figure:bilayer_pcf_rs1_z2},
respectively.  The correlation holes between the electrons in opposite wires
or planes are exceedingly shallow, although they extend over a distance
roughly equal to the separation. It can be seen that the interwire PCF has the
long-ranged oscillatory behavior exhibited by 1D HEG PCFs.

In conclusion we have used QMC to obtain the first accurate binding-energy
data for pairs of thin, parallel, metallic wires and layers.  Our results are
in broad agreement with recent RPA calculations of the binding energy and
complete disagreement with the standard pairwise vdW model.  However there are
significant differences between the DMC and RPA results for bilayers: at high
densities the asymptotic behavior of the binding energy as a function of
separation is the same but the prefactor is different, and at low densities
the DMC binding energy falls off more rapidly, implying that correlation
effects neglected in the RPA are important in this regime.  Our data can serve
as a benchmark for future theoretical studies of the binding energies of 1D
and 2D HEGs, and can also be used to parametrize model interactions between
thin conductors.  Together with our results for biwire and bilayer PCFs, our
data could be used to investigate energy functionals that incorporate vdW
effects for use in density-functional calculations.

We acknowledge financial support from Jesus College, Cambridge and the UK
Engineering and Physical Sciences Research Council.  Computing resources were
provided by the Cambridge High Performance Computing Service.


\begin{thebibliography}{99}
\bibitem{hartree_au} We use Hartree atomic units
($\hbar=|e|=m_e=4\pi\epsilon_0=1$) throughout this letter.
\bibitem{lm_comment} For pathological 1D interactions, which force the
  wave function to be 0 at points other than same-spin coalescences, the weak
  form of the Lieb-Mattis theorem implies that the energy of the paramagnetic
  phase is less than or equal to that of the ferromagnetic phase; this is
  consistent with our conclusion that the paramagnetic and ferromagnetic
  phases are degenerate in energy.
\end{thebibliography}
\end{document}